# Alternating Direction Method of Multiplier-Based Distributed Planning Model for Natural Gas, Electricity Network, and Regional Integrated Energy Systems

Ang Xuan[a], Yang Qiu[b], Yang Liu[c], Xin Sun[c],*

[a]*Tsinghua-Berkeley Shenzhen Institute, Tsinghua University, 518055 Shenzhen, China*
[b]*China Resources Power Technology Research Institute Co.Ltd, 450000 Zhengzhou, China*
[c]*Henan Electric Power Research Institite, State Grid Corporation of China, 450000 Zhengzhou, China*

**Abstract**

Regional integrated energy system coupling with multienergy devices, energy storage devices, and renewable energy devices has been regarded as one of the most promising solutions for future energy systems. Planning for existing natural gas and electricity network expansion, regional integrated energy system locations, or system equipment types and capacities are urgent problems in infrastructure development. This article employs a joint planning model to address these; however, the joint planning model ignores the potential ownerships by three agents, for which investment decisions are generally made by different investors. In this work, the joint planning model is decomposed into three distributed planning subproblems related to the corresponding stakeholders, and the alternating direction method of multipliers is adopted to solve the tripartite distributed planning problem. The effectiveness of the planning model is verified on an updated version of the Institute of Electrical and Electronics Engineers (IEEE) 24-bus electric system, the Belgian 20-node natural gas system, and three assumed integrated energy systems. Simulation results illustrate that a distributed planning model is more sensitive to individual load differences, which is precisely the defect of the joint planning model. Moreover, the algorithm performance considering rates of convergence and the impacts of penalty parameters is further analyzed.

## 1. Introduction

Faced with the global depletion of fossil energy, regional integrated energy systems (RIESs), in which electricity, gas, heating, and cooling interact with one another, have been regarded as a promising solution for future energy needs [1]. The penetration of intermittent renewable energy systems (RES) such as photovoltaics and wind turbines has received increasing attention in recent years [2]; meanwhile, energy storage systems (ESSs) are widely incorporated with RIESs due to their vital roles in alleviating fluctuations of renewable energy and load, peak shaving, ensuring uninterruptible power supply, and facilitating energy arbitrage [3]. An energy hub (EH) is a unit where transformation, conversion, and storage of various forms of energy are centralized to model RIESs [4].

---

* Corresponding author. Tel.: +86-18601110076.
*E-mail address:* sunxin3@ha.sgcc.com.cn

RIES planning problems require comprehensive consideration of all involved energy systems instead of a single energy system. There have been several works focusing on these planning problems in recent years. Zhang et al. [5] proposed a planning model for configuring an RIES for a district energy sector, and Xuan et al. [6] presented a two-stage mixed-integer linear programming approach for RIES planning based on conditional value-at-risk considering ESS and RES. These works focused on the interior structure of a single EH and ignored siting problems. Furthermore, there have been a variety of works on the joint planning of energy systems considering natural gas networks (GNs) and electricity networks (ENs). Expansion planning for integrated GNs and ENs has been presented [7][8], and various joint operation models have been widely applied to RIES planning, such as a bi-level joint planning framework [9], multistage stochastic planning framework [10], and robust joint planning framework [11].

These discussed research works are mainly aimed at problems with one or several parts of energy system expansion planning, RIES siting, or equipment configuration. Several works have focused on formulating a general joint planning problem, taking the above planning problems into account. Bent et al. [13] proposed joint expansion planning for GNs and ENs considering market conditions. Zhang et al. [13] studied security-constrained and N-1 contingency criterion joint expansion planning problems of combined GNs and ENs. A joint planning aggregated active distribution and transportation network considering traffic loads was proposed to model a fast-charging station [14].

The joint planning models mentioned above assume that GNs, ENs, and RIESs are possessed and operated by a single stakeholder, but that is not the usual situation. It is more practicable to divide these into three independent subsystems in light of actual circumstances, therefore developing a distributed optimization framework.

To address the GN-EN-RIES tripartite planning problem, the alternating direction method of multipliers (ADMM) distributed algorithm [15] is employed in this paper. ADMM brings in dual multipliers to relax equality constraints to decompose the original problem into mutually independent subproblems, namely blocks; the variables and constraints from one subproblem will not occur in others. Lin, Ma, and Zhang [16] showed that a three-block ADMM globally converges with any penalty parameter larger than zero if the three objective functions are smooth and strongly convex. Distributed operation problems based on ADMM have been studied in recent years, such as using distributed scheduling frameworks [17]–[19] to distinguish GN and EN as independent stakeholders, or a robust optimal scheduling model to study coordinated operation of energy systems via ADMM [20]. One study regarded the electricity-gas network as one stakeholder and district RIES as another to study distributed scheduling [21].

However, the existing research mainly has focused on scheduling problems and rarely involves planning aspects, and the proposed models are not universal. For this reason, this paper proposes an ADMM-based general distributed planning model for GNs, ENs, and RIESs. The main contributions are summarized as follows:

- A three-agent (GN-EN-RIES) distributed planning model is proposed considering GN/EN expansion, RIES siting, and RIES component equipment type and capacity selection, and in which the GN, EN, and RIES are owned by different stakeholders.
- An ADMM-based three-agent distributed optimization algorithm is novelly employed to address the planning model, in which energy flow in gas pipelines and electricity transmission lines are defined as continuous coupling iterative variables to converge transmitted and received power and update multipliers.
- Algorithm performance is further analyzed considering rate of convergence and impact of penalty parameters.

The rest of this paper is organized as follows: model formulations including GN/EN/RIES and their power-balancing connections are introduced in Section II, the methodology of the joint and distributed planning models are described in Section III, a case study is discussed in Section IV, and conclusions are given in Section V concludes the paper.

**Nomenclature**

**Abbreviations**
ADMM        Alternating direction method of multipliers

| CCHP | Combined cooling, heating, and power |
|---|---|
| EB | Electric bus |
| EH | Energy hub |
| GB | Gas boiler |
| GN | Gas networks |
| SOC | State of charge |
| TL | Transmission line |
| WT | Wind turbine |
| EN | Electricity networks |
| ESS | Energy storage systems |
| RES | Renewable energy systems |
| RIES | Regional integrated energy systems |

**Sets**

| $\Omega_{GS}$ | Set of gas sources |
|---|---|
| $\Omega_{GN}$ | Set of gas network nodes |
| $\Omega_{EP}$ | Set of existing gas pipelines |
| $\Omega_{CP}$ | Set of candidate gas pipelines |
| $\Omega_{GEN}$ | Set of generators |
| $\Omega_{EB}$ | Set of electric buses |
| $\Omega_{EL}$ | Set of existing electricity transmission lines |
| $\Omega_{CL}$ | Set of candidate electricity transmission lines |
| $\Omega_{RI}$ | Set of regional integrated energy systems |

*Matrixes :*

| $\mathcal{H}$ | Energy conversion matrix |
|---|---|
| $\mathcal{A}$ | Node-gas source incidence matrix |
| $\mathcal{B}$ | Node-pipeline incidence matrix |
| $\mathcal{C}$ | Bus-generator incidence matrix |
| $\mathcal{D}$ | Bus-branch incidence matrix |

*Parameters:*

| $T$ | Time horizon |
|---|---|
| $\underline{P} / \overline{P}$ | Minimum/maximum operation power |
| $RD / RU$ | Maximum ramp down/up power |
| $W$ | Weymouth coefficient |
| $\underline{\pi} / \overline{\pi}$ | Minimum/maximum nodal gas pressure |
| $seg$ | Segment of piecewise linearization |
| $\underline{GF} / \overline{GF}$ | Minimum/maximum gas flow |
| $\underline{T}^{on} / \underline{T}^{off}$ | Minimum up/down time |
| $\underline{PF} / \overline{PF}$ | Minimum/maximum power flow |
| $x$ | Reactance of electricity transmission line |
| $\underline{P}^{in}$ | Minimum input power |
| $\eta_{ess}^{ch} / \eta_{ess}^{dis}$ | Charge/Discharge efficiency of ESS |
| $\underline{SOC}_{ess} / \overline{SOC}_{ess}$ | Minimum/Maximum state of charge of ESS |
| $L^{load}$ | Load demand |

*Variables:*

| $p$ | Operation power |
|---|---|
| $GF$ | Gas flow |
| $\pi$ | Nodal gas pressure |
| $\delta$ | Continuous variable |
| $\phi$ | Binary variable indicating selected status of segment |
| $y$ | Binary/Integer variable indicating investment status |
| $u$ | Binary variable indicating operational status |
| $v$ | Binary variable indicating startup status |
| $w$ | Binary variable indicating shutdown status |
| $PF$ | Power flow |
| $\theta$ | Bus phase angle |
| $P^{in}/P^{out}$ | Input/output power of energy conversion matrix |
| $P_{ess}^{ch}/P_{ess}^{dis}$ | ESS charge/discharge power |
| $\upsilon_{ess}^{ch}/\upsilon_{ess}^{dis}$ | Binary variable indicating ESS charge/discharge status |
| $SOC_{ess}$ | ESS state of charge |
| $\eta_{ess}^{ch}/\eta_{ess}^{dis}$ | ESS charge/discharge power |
| $s$ | Binary variable indicating connection status |
| $TP$ | Transmitted power |
| $RP$ | Received power |
| $IP$ | Intermediate power |

## 2. Model Formulations

*2.1 GN model*

A typical GN model comprises natural gas sources, compressors, natural gas transmission pipelines, and gas loads.

*2.1.1 Natural gas sources*

Natural gas source production is limited by output constraints (Expression (1a)) and ramp constraints (Expression (1b)).

$$\underline{P_s} \leq p_{s,t} \leq \overline{P_s}, \forall s \in \Omega_{GS}, \forall t \in [1,T] \tag{1a}$$

$$-RD_s \leq p_{s,t} - p_{s,t-1} \leq RU_s, \forall s \in \Omega_{GS}, \forall t \in [2,T] \tag{1b}$$

*2.1.2 Natural gas transmission pipelines*

The Weymouth equation [22] (2) is adopted to express natural gas transmission flow in this model, in which the gas flow is expressed as a quadratic of nodal gas pressure. Expression (3) brings in the variable $I$ to replace nodal pressure $\pi$ to avoid a nonconvex function, and corresponding piecewise linearization formulas are listed as Expressions (4) and (5), adopted from [23] with continuous variable $\delta$ and binary variable $\phi$; the former indicating the proportion occupied in a specific segment and the latter indicating the selected status of that segment (the value of $\phi$ changing from 1 to 0 means it is selected).

$$GF_{p,t}\left|GF_{p,t}\right| = W_p^2 \left(I_{from,t}^{\ 2} - I_{to,t}^{\ 2}\right), \forall p \in \Omega_{EP}, \forall t \in [1,T] \tag{2}$$

$$I_{m,t} = \pi_{m,t}^2, \forall m \in \Omega_{GN}, \forall t \in [1,T] \tag{3a}$$

$$\underline{\pi_m} \leq \pi_{m,t} \leq \overline{\pi_m}, \forall m \in \Omega_{GN}, \forall t \in [1,T] \tag{3b}$$

$$GF_{p,t} = GF_{p,t,1} + \sum_{k=1}^{seg} \delta_{p,t,k}\left(GF_{p,t,k+1} - GF_{p,t,k}\right), \forall p \in \Omega_{EP} \cup \Omega_{CP}, \forall t \in [1,T] \tag{4a}$$

$$GF_{p,t}\left|GF_{p,t}\right| = GF_{p,t,1}\left|GF_{p,t,1}\right| + \sum_{k=1}^{segment} \delta_{p,t,k}\left(GF_{p,t,k+1}\left|GF_{p,t,k+1}\right| - GF_{p,t,k}\left|GF_{p,t,k}\right|\right), \forall p \in \Omega_{EP} \cup \Omega_{CP}, \forall t \in [1,T] \tag{4b}$$

$$0 \leq \delta_{p,t,k} \leq 1, \forall p \in \Omega_{EP} \cup \Omega_{CP}, \forall k = [1, seg], \forall t \in [1,T] \tag{5a}$$

$$\phi_{p,t,k} \leq \delta_{p,t,k}, \forall p \in \Omega_{EP} \cup \Omega_{CP}, \forall k = [1, seg-1], \forall t \in [1,T] \tag{5b}$$

$$\delta_{p,t,k+1} \leq \phi_{p,t,k}, \forall p \in \Omega_{EP} \cup \Omega_{CP}, \forall k = [1, seg-1], \forall t \in [1,T] \tag{5c}$$

Expressions (6) and (7) are gas flow constraints for existing and candidate gas pipelines, in which $M_g$ is a large positive integer in the Big-M method. Expression (6a) implies that gas flow will not meet the Weymouth equation if the compressor is installed; a similar situation of a compressor installed on a candidate pipeline and logical investment constraints are shown in Expressions (7a)–(7c).

$$-y_p^{com} M_g \leq GF_{p,t} - \left[ GF_{p,t,1} + \sum_{k=1}^{seg} \delta_{p,t,k} \left( GF_{p,t,k+1} - GF_{p,t,k} \right) \right] \leq y_p^{com} M_g, \forall p \in \Omega_{EP}, \forall t \in [1,T] \tag{6a}$$

$$-\overline{GF_p} \leq GF_{p,t} \leq \overline{GF_p}, \forall p \in \Omega_{EP}, \forall t \in [1,T] \tag{6b}$$

$$-\left[1-\left(y_p - y_p^{com}\right)\right] M_g \leq GF_{p,t} - \left[ GF_{p,t,1} + \sum_{k=1}^{seg} \delta_{p,t,k} \left( GF_{p,t,k+1} - GF_{p,t,k} \right) \right] \leq \left[1-\left(y_p - y_p^{com}\right)\right] M_g, \forall p \in \Omega_{CP}, \forall t \in [1,T] \tag{7a}$$

$$y_p^{com} \leq y_p, \forall p \in \Omega_{CP} \tag{7b}$$

$$-y_p \overline{GF_p} \leq GF_{p,t} \leq y_p \overline{GF_p}, \forall p \in \Omega_{CP}, \forall t \in [1,T] \tag{7c}$$

*2.2 EN model*

A typical electricity system comprises generators, electricity transmission lines, and electricity loads.

*2.2.1 Generators*

The production of generators is limited by output constraints (Expression (8a)) and ramp constraints (Expression (8b)).

$$u_{j,t} \underline{P_{j,t}} \leq P_{j,t} \leq u_{j,t} \overline{P_{j,t}}, \forall j \in \Omega_{GEN}, \forall t \in [1,T] \tag{8a}$$

$$-RD_j \leq P_{j,t} - P_{j,t-1} \leq RU_j, \forall j \in \Omega_{GEN}, \forall t \in [2,T] \tag{8b}$$

$$\sum_{tt=t-T_n^{on}+1}^{t} v_{j,tt} \leq u_{j,t}, \forall j \in \Omega_{GEN}, \forall t \in \left[T_j^{on}, T\right], \forall t \in [1,T] \tag{9a}$$

$$\sum_{tt=t-T_n^{off}+1}^{t} w_{j,tt} \leq 1 - u_{j,t}, \forall j \in \Omega_{GEN}, \forall t \in \left[T_j^{off}, T\right] \tag{9b}$$

$$u_{j,t} - u_{j,t-1} = v_{j,t} - w_{j,t}, \forall j \in \Omega_{GEN}, \forall t \in [2,T] \tag{9c}$$

$$v_{j,t} + w_{j,t} \leq 1, \forall j \in \Omega_{GEN}, \forall t \in [2,T] \tag{9d}$$

The unit commitment constraints in Expressions (9a)–(9d) are employed to accurately describe generator output characteristics. $v$ and $w$ are binary variables that reflect startup and shutdown actions, respectively; the value of $v$ is 1 if the generator started up from the prior moment, and the value of $w$ is 1 if the generator shut down from the prior moment, otherwise they remain 0.

*2.2.2 Electricity transmission lines*

$$PF_{l,t} = \left(\theta_{from,t} - \theta_{to,t}\right) / x_l, \forall l \in \Omega_{EL}, \forall t \in [1,T] \tag{10}$$

$$-(1-y_l)M_E \leq PF_{l,t} - \left(\theta_{from,t} - \theta_{to,t}\right) / x_l \leq (1-y_l)M_E, \forall l \in \Omega_{CL}, \forall t \in [1,T] \tag{11}$$

$$-\overline{PF_l} \leq PF_{l,t} \leq \overline{PF_l}, \forall l \in \Omega_{EL}, \forall t \in [1,T] \tag{12a}$$

$$-y_l \overline{PF_l} \leq PF_{l,t} \leq y_l \overline{PF_l}, \forall l \in \Omega_{CL}, \forall t \in [1,T] \tag{12b}$$

$$-\pi \leq \theta_{n,t} \leq \pi, \forall n \in \Omega_{EB}, \forall t \in [1,T] \tag{13}$$

The DC power flow model in equation (10) can be adopted to estimate steady state for the distribution system. $M_E$ in Expression (11) is a large positive integer in the Big-M method, set to be $2\pi / x_l$ to ensure power flow is zero if transmission line $l$ is not chosen for investment. Expressions (12)–(13) are constraints for power flow on the existing transmission line, candidate transmission line, and nodal phase, respectively.

*2.3 RIES model*

The equipment to be planned in one RIES can be divided into three types: 1) multienergy coupling devices such as combined cooling, heating, and power (CCHP); gas boiler (GB); air conditioner (AC); 2) RES such as photovoltaics (PVs) or wind turbines; or 3) ESS such as battery (BESS), thermal (TESS), and cold (CESS). An example RIES internal component configuration is shown in Fig. 1.

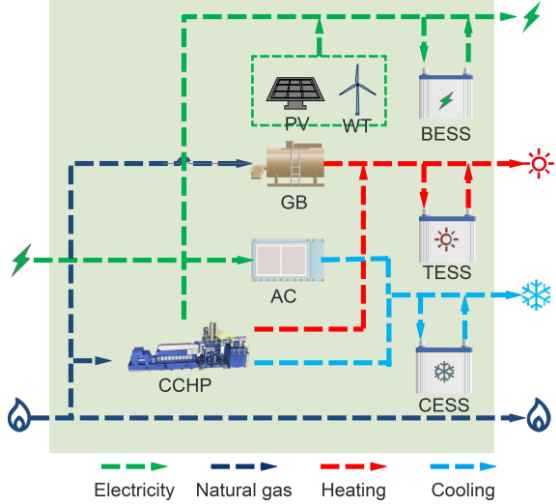

Fig. 1. Typical RIES configuration diagram

*2.3.1 Multienergy coupling device component*

$$p_{k,t}^{out} = \mathcal{H}_{k,can} p_{Can,t}^{in}, k \in \{e,h,c,g\}, \forall t \in [1,T] \tag{14}$$

$$0 \leq p_{Can,t}^{in} \leq y_{can} \overline{p_{Can,t}^{in}}, \forall can \in \{TL,TP,CCHP,GB,AC\}, \forall t \in [1,T] \tag{15}$$

The relationships between input and output power of one multienergy coupling device can be linearized in matrix form based on EH theory [4] in Expression (14), where the subscript $k$ represents energy forms including electricity, heating, cooling, and gas. Input power is constrained in Expression (15), where the candidate can be chosen from electricity transmission line, gas transmission pipeline (TP), CCHP, GB, AC, and electric bus, among others.

*2.3.2 RES component*

$$0 \leq p_{res,t} \leq y_{res} \overline{P_{res,t}^{out}}, res \in \{WT,PV\}, \forall t \in [1,T] \tag{16}$$

Renewable operation power is ruled by the maximum output power in Expression (16), where the integer decision variable $y_{res}$ indicates the number of RES component modules to be planned.

*2.3.3 ESS component*

$$0 \leq p_{ess,t}^{ch} \leq y_{ess} \overline{P_{ess}^{ch}}, \forall t \in [1,T] \tag{17a}$$

$$0 \leq p_{ess,t}^{dis} \leq y_{ess} \overline{P_{ess}^{dis}}, \forall t \in [1,T] \tag{17b}$$

$$0 \leq p_{ess,t}^{ch} \leq \upsilon_{ess,t}^{ch} M, \forall t \in [1,T] \tag{18a}$$

$$0 \leq p_{ess,t}^{dis} \leq \upsilon_{ess,t}^{dis} M, \forall t \in [1,T] \tag{18b}$$

$$\upsilon_{ess,t}^{ch} + \upsilon_{ess,t}^{dis} \leq 1, \forall t \in [1,T] \quad (19)$$

$$SOC_{ess,t} = SOC_{ess,t-1} + \eta_{ess}^{ch} P_{ess,t}^{in} - P_{ess,t}^{dis}/\eta_{ess}^{dis}, \forall t \in [1,T] \quad (20a)$$

$$y_{ess}\underline{SOC_{ess}} \leq SOC_{ess,t} \leq y_{ess}\overline{SOC_{ess}}, \forall t \in [1,T] \quad (20b)$$

$$SOC_{ess,1} = SOC_{ess,T}, ess \in \{BESS, TESS, CESS\} \quad (20c)$$

ESS charging/discharging power is limited by the power of investment options in Expressions (17)–(18), in which the upper bound is denoted by the number of ESS modules $y_{ESS}$. The binary variables $v_{s,t}^{ch,n}/v_{s,t}^{dis,n}$ denote the charge/discharge states, while $M$ is a large number used in the Big-M method. The constraint in Expression (19) ensures that charge and discharge behavior cannot happen at the same time. Expression (20) represents the state of charge calculation formula and constraints.

*2.3.4 RIES Load power balance*

$$P_{e,t}^{out} - P_{BESS,t}^{ch} + P_{BESS,t}^{dis} + P_{WT,t} + P_{PV,t} = L_{e,t}^{load}, \forall t \in [1,T] \quad (21a)$$

$$P_{h,t}^{out} - P_{TESS,t}^{ch} + P_{TESS,t}^{dis} = L_{h,t}^{load}, \forall t \in [1,T] \quad (21b)$$

$$P_{c,t}^{out} - P_{CESS,t}^{ch} + P_{CESS,t}^{dis} = L_{c,t}^{load}, \forall t \in [1,T] \quad (21c)$$

$$P_{g,t}^{out} = L_{g,t}^{load}, \forall t \in [1,T] \quad (21d)$$

Expressions (21a)–(21d) are the power balance constraints on the supply side.

*2.3.5 RIES siting*

$$\sum_h s_{m,h} \geq 1, \forall m \in \Omega_{GN}, \forall h \in \Omega_{RI} \quad (22)$$

$$\sum_h s_{n,h} \geq 1, \forall n \in \Omega_{EN}, \forall h \in \Omega_{RI} \quad (23)$$

Expressions (22) and (23) mean that every RIES can only be connected to one node/bus in a GN/EN due to investment logic. $s_{m,h}, s_{n,h}$ are binary variables indicating the connection status between GN/EN and RIES, which have a value of 1 if they are connected; otherwise they are 0.

*2.4 Power-balancing connection model*

$$0 \leq TP_{m,h,t}^g/IP_{m,h,t}^g/RP_{m,h,t}^g \leq s_{m,h}M, \forall m \in \Omega_{GN}, \forall h \in \Omega_{RI}, \forall t \in [1,T] \quad (24)$$

$$0 \leq TP_{n,h,t}^e/IP_{n,h,t}^e/RP_{n,h,t}^e \leq s_{n,h}M, \forall n \in \Omega_{EN}, \forall h \in \Omega_{RI}, \forall t \in [1,T] \quad (25)$$

$IP$ is the intermediate power between the transmitted power $TP$ and received power $RP$, its value is usually greater than 0 due to the geographical distance in the actual situation; therefore, transmitting and receiving are not completed at the same time. $s_{m,h}, s_{n,h}$ bundle the transmitted power, intermediate power, and received power using the Big-M method in the transmission procedure.

$$\mathcal{A}_{m,s}P_{s,t} + \sum_{p \in \Omega_{EP}} \mathcal{B}_{m,p}GF_{p,t} + \sum_{p \in \Omega_{CP}} \mathcal{B}_{m,p}GF_{p,t} = \sum_h TP_{m,h,t}^g + L_{m,t}^{load}, \forall m \in \Omega_{GN}, \forall s \in \Omega_{GS}, \forall h \in \Omega_{RI}, \forall t \in [1,T] \quad (26)$$

$$\mathcal{C}_{n,j}P_{j,t} + \sum_{l \in \Omega_{EL}} \mathcal{D}_{n,l}PF_{l,t} + \sum_{l \in \Omega_{CL}} \mathcal{D}_{n,l}PF_{l,t} = \sum_h TP_{n,h,t}^e + L_{n,t}^{load}, \forall n \in \Omega_{EN}, \forall j \in \Omega_{GEN}, \forall h \in \Omega_{RI}, \forall t \in [1,T] \quad (27)$$

$$\sum_m RP_{m,h,t}^g = P_{h,CCHP,t} + P_{h,GB,t} + P_{h,TP,t}, \forall m \in \Omega_{GN}, \forall h \in \Omega_{RI}, \forall t \in [1,T] \quad (28a)$$

$$\sum_n RP_{n,h,t}^e = P_{h,TL,t} + P_{h,AC,t}, \forall n \in \Omega_{EN}, \forall h \in \Omega_{RI}, \forall t \in [1,T] \quad (28b)$$

The power balances at the GN-EN-RIES tripartite interfaces can be modeled as Expressions (26)–(28). Expression (26) is the natural gas balance among generator, existing pipeline, candidate pipeline, transmitted gas flow, and load, where $\mathcal{A}_{m,s}$ equals 1 if gas source $s$ is connected to node $m$, $\mathcal{B}_{m,p}$ equals 1 if pipeline $p$ starts at node $m$, $\mathcal{B}_{m,p}$ equals −1 if pipeline $p$ ends at node $m$, otherwise their values are 0. Expression (27) is the bus electricity balance function, where $\mathcal{C}_{n,j}$ equals 1 if generator $j$ is connected to bus $n$, $\mathcal{D}_{n,l}$ equals 1 if branch $l$ starts at bus $n$, $\mathcal{D}_{n,l}$ equals $-1$ if branch $l$ ends at bus $n$, otherwise their values are 0. Expressions (28a) and (28b) are power balance constraints on the input side for a single RIES.

## 3. Methodology

### 3.1 Joint planning model

The centralized joint planning model can be formulated as follows.

$$\min \left\{ \begin{array}{l} \sum_{p \in \Omega_{CP}} c_p y_p + \sum_{com \in \Omega_{COM}} c_{com} y_{com} + \sum_{l \in \Omega_{CL}} c_l y_l + \sum_t \left( \sum_{m \in \Omega_{GN}} c_m P_{m,t} + \sum_{n \in \Omega_{EN}} c_n P_{n,t} \right) \\ + \sum_{h \in \Omega_{RI}} \left( \sum_{can} c_{can} y_{h,can} + \sum_{res} c_{res} y_{h,res} + \sum_{ess} c_{ess} y_{h,ess} \right) \\ + \sum_{h \in \Omega_{RI}} \left( \sum_{m \in \Omega_{GN}} c_m s_{m,h} + \sum_{n \in \Omega_{EN}} c_n s_{n,h} \right) \\ + \sum_t \sum_{h \in \Omega_{RI}} \left( Pr^g \sum_{m \in \Omega_{GN}} RP^g_{m,h,t} + Pr^e \sum_{n \in \Omega_{EN}} RP^e_{n,h,t} \right) \end{array} \right\} \quad (29)$$

$$\begin{array}{l} TP^g_{m,h,t} = RP^g_{m,h,t}, \forall m \in \Omega_{GN}, \forall h \in \Omega_{RI}, \forall t \in [1,T] \\ TP^e_{n,h,t} = RP^e_{n,h,t}, \forall n \in \Omega_{EN}, \forall h \in \Omega_{RI}, \forall t \in [1,T] \end{array} \quad (30)$$

s.t. (1a)–(28b)

The objective function in Expression (29) minimizes the total investment costs and trading costs for GN, EN, and RIES. The first three terms represent the investment costs of candidate transmission pipelines, pipelines, and compressors, the fourth term represents operational costs of gas sources and generators, the next two terms represent investment costs of RIES components and siting cost, and the last term represents the trading costs to the RIES of purchasing electricity and gas from external networks. Expression (30) ensures the equality of transmitted and received power in the same channel; the formula connects three separate planning agents directly and ignores the existence of intermediate power, namely, the default is that power flow transmission and reception are completed simultaneously.

### 3.2 Background on ADMM optimization theory

ADMM is an algorithm that solves convex optimization problems by breaking them into smaller blocks, each of which is then easier to handle. It has found wide applications in a large number of areas with several implementations [17]-[21]. It is impossible to illustrate these applications comprehensively here; we only intend to highlight the primary unscaled form of ADMM as follows (based on [15]). Expression (31) is a basic optimization problem, in which $f/g$ are separable convex objective functions for two sets of variables $x/z$. The Augmented Lagrangian Function (ALF) can be acquired by adding a squared penalty term (coefficient $\rho/2$) to the primary Lagrangian function in Expression (32).

$$\min_{x,z} f(x) + g(z) \quad (31)$$

$$s.t. Ax + Bz \leq c$$

$$L_\rho(x,z,\lambda) = f(x) + g(z) + \lambda^T (Ax + Bz - c) + \frac{\rho}{2} \|Ax + Bz - c\|_2^2 \quad (32)$$

$$\left(x^{k+1}, z^{k+1}\right) = \arg\min_{x,z} L_\rho(x,z,\lambda^k) \quad (33)$$

$$x^{k+1} = \arg\min_x L_\rho(x, z^k, \lambda^k) \quad (34a)$$

$$z^{k+1} = \arg\min_{z} L_\rho(x^{k+1}, z, \lambda^k) \tag{34b}$$

$$\lambda^{k+1} = \lambda^k + \rho(Ax^{k+1} + Bz^{k+1} - c) \tag{34c}$$

The ADMM is changed to alternately iterate $x, z$ alone (Expressions (34a)–(34c)) based on the original foundation in Expression (33) ($x, z$ iterate together). The updated formula of the alternative multiplier $\lambda$ is listed in Expression (34c). The novel application of the ADMM algorithm to GN-EN-RIES planning in this work has expanded three-block mathematical optimization to a three-agent engineering model.

*3.3 GN-EN-RIES three-agent distributed planning model*

The joint planning model ignores the different ownerships of the tripartite agents and assumes that *TP* does not exist all the time (Expression (30)), while distributed planning considers a more practical model. Different from the objective function in Expression (29), ADMM-based distributed planning decomposes the original objective function to three ALFs with squared penalty terms, given in Expressions (35) to (37). The distributed planning model framework is listed in Table , by which the three agents can be optimized to achieve their own optimal investment decisions separately.

$$\min L_{GN} = \min \left\{ \begin{array}{l} \sum_{p \in \Omega_{CP}} c_p y_p + \sum_{com \in \Omega_{COM}} c_{com} y_{com} + \sum_t \sum_{m \in \Omega_{GN}} c_m P_{m,t} \\ + \sum_t \sum_{h \in \Omega_{RI}} \sum_{m \in \Omega_{GN}} \lambda_{m,h,t}^{gn} \left( TP_{m,h,t}^g - IP_{m,h,t}^g \right) \\ + \dfrac{\rho^{gn}}{2} \sum_t \sum_{h \in \Omega_{RI}} \sum_{m \in \Omega_{GN}} \left( TP_{m,h,t}^g - IP_{m,h,t}^g \right)^2 \end{array} \right\} \tag{35}$$

$$\min L_{EN} = \min \left\{ \begin{array}{l} \sum_{l \in \Omega_{CL}} c_l y_l + \sum_t \sum_{n \in \Omega_{EN}} c_n P_{n,t} \\ + \sum_t \sum_{h \in \Omega_{RI}} \sum_{n \in \Omega_{EN}} \lambda_{n,h,t}^{en} \left( TP_{n,h,t}^e - IP_{n,h,t}^e \right) \\ + \dfrac{\rho^{en}}{2} \sum_t \sum_{h \in \Omega_{RI}} \sum_{n \in \Omega_{EN}} \left( TP_{n,h,t}^e - IP_{n,h,t}^e \right)^2 \end{array} \right\} \tag{36}$$

$$\min L_{RIES} = \min \left\{ \begin{array}{l} \sum_{h \in \Omega_{RI}} \left( \sum_{can} c_{can} y_{h,can} + \sum_{res} c_{res} y_{h,res} + \sum_{ess} c_{ess} y_{h,ess} \right) \\ + \sum_{h \in \Omega_{RI}} \left( \sum_{m \in \Omega_{GN}} c_m s_{m,h} + \sum_{n \in \Omega_{EN}} c_n s_{n,h} \right) + \sum_t \sum_{h \in \Omega_{RI}} \left( c_{TP} P_{h,TP,t} + c_{TL} P_{h,TL,t} \right) \\ + \sum_t \sum_{h \in \Omega_{RI}} \sum_{m \in \Omega_{GN}} \lambda_{m,h,t}^{ehg} \left( RP_{m,h,t}^g - IP_{m,h,t}^g \right) + \dfrac{\rho^{ehg}}{2} \sum_t \sum_{h \in \Omega_{RI}} \sum_{m \in \Omega_{GN}} \left( RP_{m,h,t}^g - IP_{m,h,t}^g \right)^2 \\ + \sum_t \sum_{h \in \Omega_{RI}} \sum_{n \in \Omega_{EN}} \lambda_{n,h,t}^{ehe} \left( RP_{n,h,t}^e - IP_{n,h,t}^e \right) + \dfrac{\rho^{ehe}}{2} \sum_t \sum_{h \in \Omega_{RI}} \sum_{n \in \Omega_{EN}} \left( RP_{n,h,t}^e - IP_{n,h,t}^e \right)^2 \end{array} \right\} \tag{37}$$

Table 1. ADMM-based GN-EN-RIES three-agent distributed planning model framework

| Agent | GN | EN | RIES |
|---|---|---|---|
| Objective Function | (35) | (36) | (37) |
| Power Balance Constraints | s.t. (26) | s.t. (27) | s.t. (28a)-(28b) |
| Other Constraints | s.t. (1)-(7b) | s.t. (8a)-(13) | s.t. (14)-(25) |

The detailed process of the proposed ADMM-based GN-EN-RIES three-agent distributed optimization algorithm is shown in Fig. 2. The iterative variables in GN-EN-RIES apply to the entire time horizon, in which the superscript $k$ represents the $k$th iteration. Any object that is not set as text should be labeled as a figure or table.

**Algorithm 1** ADMM BASED GN-EN-RIES 3-AGENT DISTRIBUTED OPTIMIZATION

1: initialize alternating multipliers $\lambda_{m,h,t}^{gn,0}, \lambda_{n,h,t}^{en,0}, \lambda_{m,h,t}^{ehg,0}, \lambda_{n,h,t}^{ehe,0}$;
2: initialize immediate variables $IP_{m,h,t}^{g,0}, IP_{n,h,t}^{e,0}$;
3: define penalty coefficient $\rho^{gn}, \rho^{en}, \rho^{ehg}, \rho^{ehe}$;
4: define convergence threshold $\epsilon^g, \epsilon^e$.
5: **for** $k=0; k++$ **do**
6: $\quad TP_{m,h,t}^{g,k+1} = \mathbf{argmin} L_{GN}(TP_{m,h,t}^g, IP_{m,h,t}^{g,k}, \lambda_{m,h,t}^{gn,k})$;
7: $\quad TP_{n,h,t}^{e,k+1} = \mathbf{argmin} L_{EN}(TP_{n,h,t}^e, IP_{n,h,t}^{e,k}, \lambda_{n,h,t}^{en,k})$;
8: $\quad (RP_{m,h,t}^{g,k+1}, RP_{n,h,t}^{e,k+1}) = \mathbf{argmin} L_{RIES}(RP_{m,h,t}^g, RP_{n,h,t}^e, IP_{m,h,t}^{g,k}, IP_{n,h,t}^{e,k}, \lambda_{m,h,t}^{ehg,k}, \lambda_{n,h,t}^{ehe,k})$;
9: $\quad$ **if** $max(| TP_{m,h,t}^{g,k+1} - IP_{m,h,t}^{g,k} |, | RP_{m,h,t}^{g,k+1} - IP_{m,h,t}^{g,k} |) \leq \epsilon^g$ &
10: $\quad\quad max(| TP_{n,h,t}^{e,k+1} - IP_{n,h,t}^{e,k} |, | RP_{n,h,t}^{e,k+1} - IP_{n,h,t}^{e,k} |) \leq \epsilon^e$ **then**
11: $\quad\quad$ **return**
12: $\quad$ **end if**
13: $\quad IP_{m,h,t}^{g,k+1} = (TP_{m,h,t}^{g,k+1} + RP_{m,h,t}^{g,k+1})/2$;
14: $\quad IP_{n,h,t}^{e,k+1} = (TP_{n,h,t}^{e,k+1} + RP_{n,h,t}^{e,k+1})/2$;
15: $\quad \lambda_{m,h,t}^{gn,k+1} = \lambda_{m,h,t}^{gn,k} + \rho^{gn}(TP_{m,h,t}^{g,k+1} - IP_{m,h,t}^{g,k+1})$;
16: $\quad \lambda_{n,h,t}^{en,k+1} = \lambda_{n,h,t}^{en,k} + \rho^{en}(TP_{n,h,t}^{e,k+1} - IP_{n,h,t}^{e,k+1})$;
17: $\quad \lambda_{m,h,t}^{ehg,k+1} = \lambda_{m,h,t}^{ehg,k} + \rho^{ehg}(RP_{m,h,t}^{g,k+1} - IP_{m,h,t}^{g,k+1})$;
18: $\quad \lambda_{n,h,t}^{ehe,k+1} = \lambda_{n,h,t}^{ehe,k} + \rho^{ehe}(RP_{n,h,t}^{g,k+1} - IP_{n,h,t}^{e,k+1})$;
19: **end for**

Fig. 2. Steps in the ADMM-based GN-EN-RIES three-agent distributed optimization algorithm

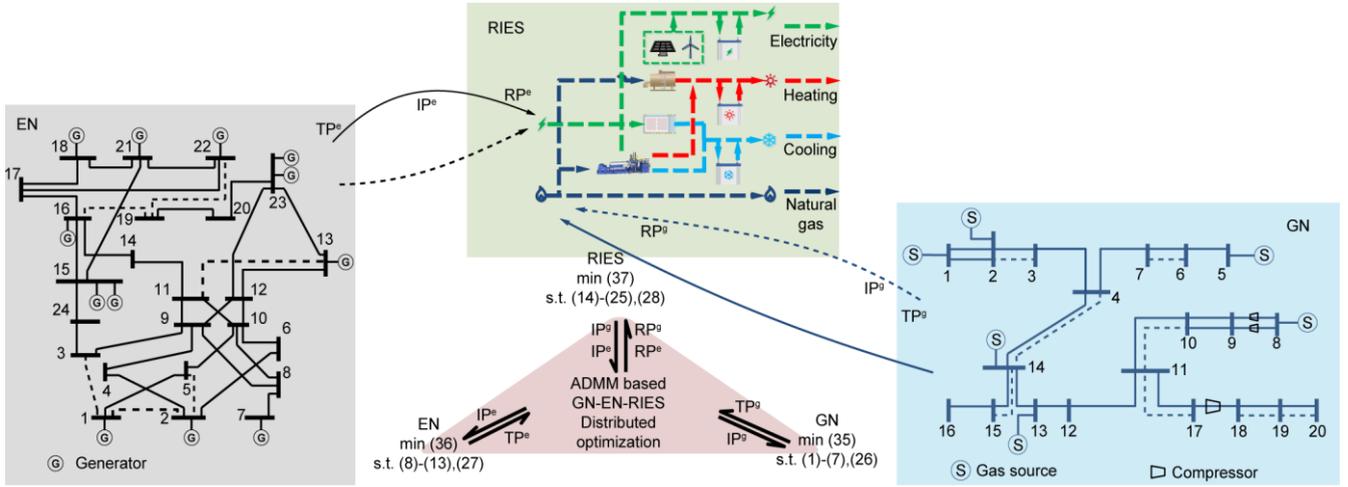

Fig. 3. Planned model composed of updated IEEE 24-bus electric system, Belgian 20-node natural gas system, and assumed RIES

## 4. Case Study

An updated Institute of Electrical and Electronics Engineers (IEEE) 24-bus electric system [24], the Belgian 20-node natural gas system [25], and three assumed RIESs (Appendix A. ) are employed to verify the effectiveness of the proposed joint and distributed models; all the datasets are available in [26]. The large mixed-integer programming (MIP) is modeled on Matlab 2019b with YALMIP and solved by Gurobi 9.1 on a Lenovo laptop with Intel® Core™ i7-6700U 3.40 GHz processor and 8GB RAM. $\lambda_{m,h,t}^{gn}, \lambda_{n,h,t}^{en}, \lambda_{m,h,t}^{ehg}, \lambda_{n,h,t}^{ehe}$ are initialized to 1, $IP_{m,h,t}^{g}, IP_{n,h,t}^{e}$ are initialized to 0, $\rho^{gn}, \rho^{en}, \rho^{ehg}, \rho^{ehe}$ are set to 100, and $\varepsilon^{g}, \varepsilon^{e}$ are set to $10^{-3}$. The gas flow in the GN piecewise linearization is set as three segments, and the prices of purchasing electricity and gas in the RIES are 0.3 US\$/KWh and 0.5 US\$/m$^3$.

To verify the effectiveness and increase diversity, three district RIESs are assumed to represent different load conditions, in which one is designed without cold load, one is designed without heat load, and the third is designed as a mix between the other two. A sketch map of GN/EN expansion planning and RIES siting planning is shown in Fig. , in which one RIES is chosen to simplify the expression. The different colored backgrounds represent the different ownerships of corresponding stakeholders.

*4.1 GN-EN Expansion and RIES site selection results*

The detailed planning results obtained for the two models are displayed in Table 2. Planning results illustrate that the distributed model requires more investment than the joint model. But it cannot be concluded that the joint model is better than the proposed distributed one; on the contrary, the joint planning model ignores energy flows in the transmission process, which is an important part of actual engineering problems. GN-EN-RIES are usually geographically distant from one another, and power flow and gas flow exist in the transmission process, thus, ignoring them will cause large planning errors. Moreover, if the investor realizes the functional expression of energy flow, it can be added as a new constraint to $IP$, therefore making the modeling result more consistent with the actual situation.

Table 2. Expansion and siting planning results obtained for the two models

|  |  | Joint planning | Distributed planning |
|---|---|---|---|
| Expansion | GN | Pipelines: None Compressors: None | Pipelines: None Compressors: None |
|  | EN | 19–22 | 1–3, 11–13, 19–22 |
| Siting | RIES1 | GN: 17 EB: 3 | GN: 17 EB: 3 |
|  | RIES2 | GN: 17 EB: 3 | GN: 17 EB: ALL |
|  | RIES3 | GN: 17 EB: 3 | GN: 17 EB: ALL |
| Total Cost (M$) |  | 42.19 | 47.62 (+12.87%) |

*GN = Gas Node, EB = Electric Bus

In terms of the planning costs of the two approaches, a single stakeholder is assumed to own the whole GN-EN-RIES in the joint planning model, while the GN-EN-RIES has three separate ownerships in the distributed model. Moreover, convergence means that each stakeholder achieves consensus about transmission power with acceptable accuracy $\varepsilon^g, \varepsilon^e$ through the iteration process. By contrast, the total cost of distributed planning is 12.87% larger than the global economic optimal solution. More details on computational inputs are given in Table 3.

Table 3. Investment and operation costs obtained for the two models

| Objective (M$) |  | Joint planning | Distributed planning |
|---|---|---|---|
| Investment Cost (M$) | GN pipeline | 0 | 0 |
|  | GN compressor | 0 | 0 |
|  | EN transmission line | 0.05 | 0.15 |
|  | RIES component | 8.63 | 9.68 |
|  | RIES siting | 0.42 | 5.50 |
| Operation Cost (M$) | GN | 12.85 | 12.85 |
|  | EN | 0.43 | 0.42 |
|  | RIES purchased electricity | 19.78 | 19.00 |
|  | RIES purchased gas | 0 | 0 |
| Total Cost (M$) |  | 42.19 | 47.62 (+12.87%) |

*4.2 RIES component planning results*

RIES configuration planning results are further analyzed in Table 4. Generally speaking, the joint planning model has less investment due to the goal of global economic optimization (given by the objective function in Expression (29)). As for the distributed model, it has a more significant impact on site planning due to the hysteresis effect of transmission energy flow. The actual energy supplied by outer pipeline/transmission lines is usually smaller than transmitted energy, thus more investment in transmission is required.

Table 4. RIES component planning results

| | | Joint planning | Distributed planning |
|---|---|---|---|
| RIES1 | CCHP (MW) | 1 | 1 |
| | GB (MW) | 6 | 6 |
| | AC (MW) | 0 | 0 |
| | BESS (MWh) | 0 | 0 |
| | TESS (MWh) | 2.32 | 2.32 |
| | CESS (MWh) | 0 | 0 |
| | WT (MW) | 0.2 | 0.2 |
| | PV (MW) | 0.05 | 0.05 |
| RIES2 | CCHP (MW) | 1 | 1 |
| | GB (MW) | 0 | 0 |
| | AC (MW) | 6 | 6 |
| | BESS (MWh) | 0 | 0 |
| | TESS (MWh) | 0 | 0 |
| | CESS (MWh) | 8 | 8 |
| | WT (MW) | 0.2 | 0.2 |
| | PV (MW) | 0.05 | 0.05 |
| RIES3 | CCHP (MW) | 2 | 3 |
| | GB (MW) | 1 | 0 |
| | AC (MW) | 1 | 1 |
| | BESS (MWh) | 0 | 0 |
| | TESS (MWh) | 1.16 | 2.64 |
| | CESS (MWh) | 2.22 | 1.33 |
| | WT (MW) | 0.15 | 0.13 |
| | PV (MW) | 0.03 | 0.03 |

*4.3 Algorithm performance analysis*

Although a rigorous convergence proof is beyond the scope of this work, the algorithm performance of the proposed ADMM application in power systems is still demonstrated. Rates of convergence and the impacts of penalty parameters are analyzed further.

*1) Rate of convergence*

The time consumption of the original joint algorithm is 139.71 s, with a faster calculation speed compared to the proposed distributed optimization algorithm (in Table 5). The proposed algorithm has a convergence time of 2000 s for the convergence thresholds $\varepsilon^g, \varepsilon^e$ of $10^{-3}$, and the number of iteration steps is 13 (in **Error! Reference source not found.**.(d)) due to the interactions among power flows. However, within the engineering scope for the GN, EN, and RIES planning problem the indices are completely acceptable. If the convergence threshold ($10^{-2}$ or larger) and gap tolerance of the MIP solver Gurobi ($10^{-6}$ or larger) are appropriately increased, calculation times could be further improved.

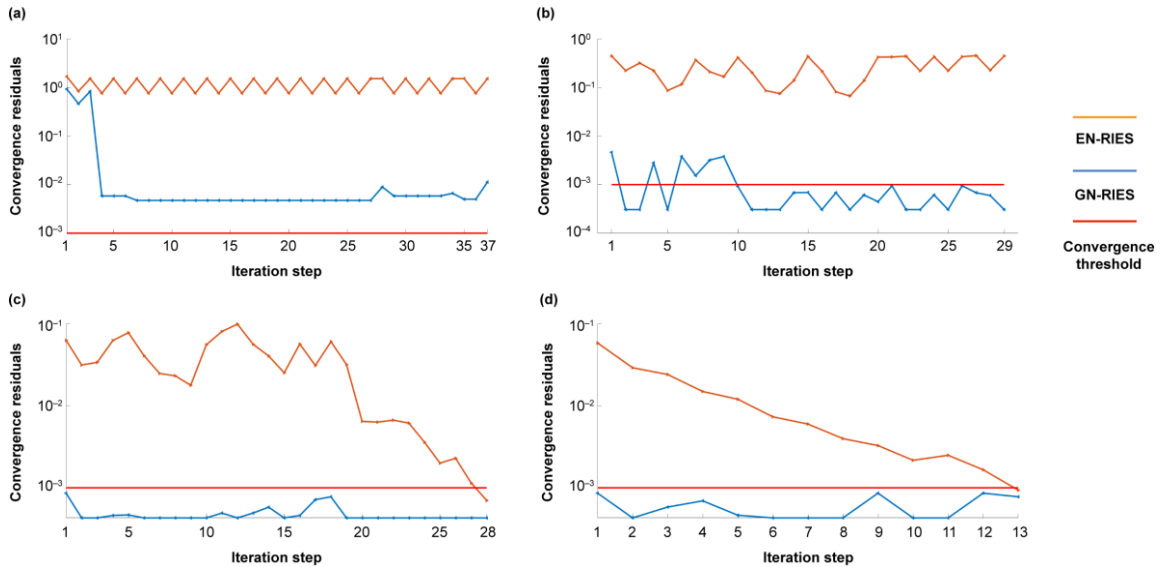

Fig. 4. Convergence residual curves: (a) $\rho^{gn}, \rho^{en}, \rho^{ehg}, \rho^{ehe} = (0.1, 0.1, 0.1, 0.1)$; (b) $\rho^{gn}, \rho^{en}, \rho^{ehg}, \rho^{ehe} = (1,1,1,1)$; (c) $\rho^{gn}, \rho^{en}, \rho^{ehg}, \rho^{ehe} = (10,10,10,10)$; (d) $\rho^{gn}, \rho^{en}, \rho^{ehg}, \rho^{ehe} = (100,100,100,100)$.

To further analyze the convergence performance of the proposed algorithm, the change curves of convergence residuals $\max(|TP_{n,h,t}^{e,k+1} - IP_{n,h,t}^{g,k}|, |RP_{n,h,t}^{g,k+1} - IP_{n,h,t}^{e,k}|)$ and $\max(|TP_{n,h,t}^{e,k+1} - IP_{n,h,t}^{g,k}|, |RP_{n,h,t}^{g,k+1} - IP_{n,h,t}^{e,k}|)$ in the GN-RIES system and EN-RIES system through the iteration process are shown in **Error! Reference source not found.**. The convergence curves resolve quickly, but it should be noted that the residuals of the GN-RIES system (blue line) reached convergence threshold first, but the corresponding parameters of the EN-RIES system (orange line) have difficulty meeting convergence. This is because the gas load assumed is quite small and there is no need to purchase a large amount of natural gas for RIESs through the gas pipeline network, thus, it is easier to converge with $IP$ ($IP$ is initialized to 0). Only the residuals from the GN-RIES system and the EN-RIES system need to satisfy convergence accuracy before the system can be judged to meet the convergence requirements.

*2) The impact of penalty parameters*

The values of penalty parameters, also known as convergence factors in *Operation Research*, were changed in multiple simulation calculations to observe the impacts on convergence performance. The different numbers of iteration steps and times to convergence resulting from the penalty parameters are listed in Table 5. The following

Table 5. Algorithm performance under different penalty parameters

| Penalty parameters $\rho^{gn}, \rho^{en}, \rho^{ehg}, \rho^{ehe}$ | Result | Iteration step | Convergence time (s) |
|---|---|---|---|
| (0.1,0.1,0.1,0.1) | - | - | >20000 |
| (1,1,1,1) | - | - | >20000 |
| (10,10,10,10) | 47.68 | 28 | 5644.65 |
| (100,100,100,100) | 47.62 | 13 | 1780.23 |
| (200,200,200,200) | 47.66 | 17 | 2383,47 |
| (10,10,100,100) | 47.64 | 24 | 4276.34 |
| (100,10,100,100) | 47.64 | 24 | 4281.15 |
| (10,100,100,100) | 47.58 | 14 | 1739.32 |
| (100,100,10,10) | 47.64 | 26 | 5066.90 |

- No convergence

conclusions can be obtained:

Variations of the penalty parameters $\rho^{gn}, \rho^{en}, \rho^{ehg}, \rho^{ehe}$ will affect convergence accuracy and speed of the algorithm, shown in **Error! Reference source not found.** and Table 5. We set $\rho^{gn}, \rho^{en}, \rho^{ehg}, \rho^{ehe}$ to different values to analyze the trend, and when the value of the penalty parameter is less than 1, the calculation process cannot converge. The dual residuals show fluctuations with iterations in Fig. (a)(b), and the algorithm cannot converge when the values of the penalty parameters are 1 or smaller; when the penalty parameter is greater than 10, the problem can gradually converge. Meanwhile, the iteration steps and time to convergence decrease rapidly with the increase of the penalty parameter. However, when the value of the penalty parameter is larger than 200, the number of iteration steps increases again.

When the penalty parameters $\rho^{gn}, \rho^{en}, \rho^{ehg}, \rho^{ehe}$ are set to different values, the results varied widely from line 6 to line 9 in Table 5, when $\rho^{en}, \rho^{ehg}, \rho^{ehe}$ are given more weight, the iteration steps and convergence time improved greatly; when $\rho^{gn}$ is given more weight, almost no change occurred in iteration step and time to convergence.

The reasonable explanation for the above conclusions is: The penalty parameter adjusts the step speed between each iteration. When the value of the penalty parameter increases, the proportion of the penalty item increases to achieve the effect of accelerating convergence; at the same time it may cause the step to dismiss the convergence result, so that when the penalty parameter is too small, it will be slow to converge. Theoretically, the size of the penalty parameter does not change the convergence result. It is recommended that a small penalty parameter be adopted in practical problems and then gradually increased. Because of the data characteristics of the case, the EN and RIES systems take more computational time; thus, multiplying a larger penalty factor will speed up the convergence process. Therefore, in actual planning problems, one can appropriately increase the corresponding penalty parameter by observing the computational time of each subproblem, and thereby relieving the calculation burden. Overall, considering the algorithm's convergence accuracy and speed, the preferable value ranges of the penalty parameters are approximately [100, 200].

## 5. Conclusions

This work proposes a joint and an ADMM-based distributed planning model considering GN and EN expansions as well as RIES siting and configuration. To sum up, the proposed joint planning and ADMM-based distributed planning models ensure power supplies of electricity and gas, as well as heat and cold loads. The existence of energy flows in the transmission process is considered, which has stronger practical significance compared to the joint planning model. The case study verified the effectiveness of the proposed model by comparing the planning results obtained by distributed planning and the global optimum obtained by joint planning. The planning results illustrate that the tripartite blocks can be owned by three separate stakeholders to achieve decision independence through transmission power convergence, and the proposed model is rational and novel. Concurrently, the case study results proved the feasibility of the ADMM-based GN-EN-RIES three-agent distributed optimization algorithm. Moreover, the convergence rate and sensitivity of penalty parameters were studied to improve algorithm performance.

It is worth pointing out that there are still many limitations and shortcomings in this research work. For example, we adopted the piecewise linearization method to replace the Weymouth function, and the RIES load data are assumed; additionally, the values of $\lambda_{m,h,t}^{gn}, \lambda_{n,h,t}^{en}, \lambda_{m,h,t}^{ehg}, \lambda_{n,h,t}^{ehe}$, and $IP_{m,h,t}^{g}, IP_{n,h,t}^{e}$ (in Section 4) are initialized arbitrarily. Exploring the impact of these simplifications and assumptions is one of the focuses of future work. The scope and direction of future research will extend the proposed distributed planning model to complex industrial application systems, such as integrated systems spread over larger geographical areas and multiple interconnected electricity and gas systems.

## CRediT authorship contribution statement

**Ang Xuan:** Conceptualization, Methodology, Visualization, Data curation, Writing-original draft. **Yang Qiu:** , Validation, Writing-review & editing. **Yang Liu:** Investigation, Project administration. **Xin Sun:** Investigation, Project administration, Supervision.

## Acknowledgments

This work is supported by the State Grid Henan electric power company science and technology project of China (52170220009V).

## Declaration of Competing Interest

The authors declare that they have no known competing financial interests or personal relationships that could have appeared to influence the work reported in this paper.

## Appendix A. Load data for RIESs

The load types of assumed RIESs in this case are shown as Fig. , all data has been organized in Excel format in [26] for your convenience.

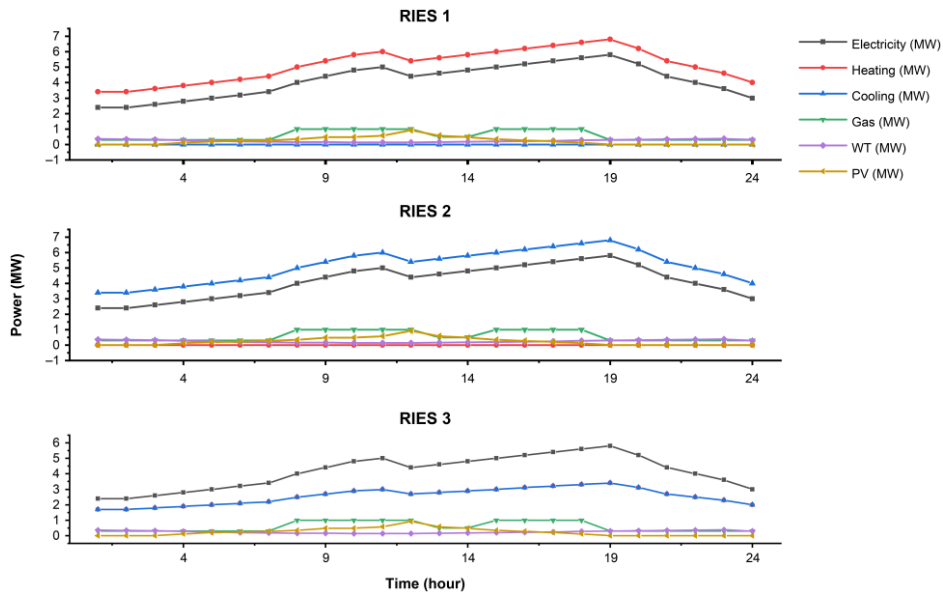

Fig. A1. Three assumed RIESs